\documentclass[aps,superscriptaddress,twocolumn,pre,longbibliography,floatfix]{revtex4-1}
\usepackage[utf8]{inputenc}
\usepackage[T1]{fontenc}
\usepackage{amssymb}
\usepackage{svg}
\usepackage[colorlinks={true}, citecolor={blue}, filecolor={blue}, linkcolor={blue}, urlcolor={blue}]{hyperref}
\usepackage{physics}
\usepackage[caption=false]{subfig}
\usepackage{graphicx}
\usepackage{soul}
\usepackage{url}
\usepackage{amsmath}
\renewcommand{\abs}[1]{\left\vert#1\right\vert}

\allowdisplaybreaks

\begin{document}
\title{Spin-1 quantum annealing with anisotropy-controlled intermediate-state pathways}
	\author{M. Haider Akbar}
 \email{makbar25@ku.edu.tr}
\affiliation{Department of Computational Sciences and Engineering, Koç University, 34450 Sarıyer, Istanbul, Türkiye}
\author{Özgür E. Müstecaplıoğlu}	
	\email{omustecap@ku.edu.tr}
	\affiliation{Department of Physics, Koç University, 34450 Sarıyer, Istanbul, Türkiye}
	\affiliation{TÜBİTAK Research Institute for Fundamental Sciences (TBAE), 41470 Gebze, Türkiye}

 \date{\today} 
\begin{abstract}
We study finite-time quantum annealing in spin-1 Ising chains with single-ion anisotropy, using clean chains to resolve the underlying dynamics and disordered ensembles with exactly known ground states to test its robustness. The anisotropy term $D\sum_i (S_i^z)^2$ acts simultaneously as an on-site localization term in the ternary cost function and as a spectral control parameter of the anneal, so that varying $D$ defines a parameterized family of ternary optimization problems. Every member of the family is benchmarked against two structurally different classical solvers acting on the identical cost function. The anisotropy controls the energetic accessibility of the intermediate $m_z = 0$ level and qualitatively restructures the annealing spectrum: the single mid-schedule avoided crossing of the easy-axis sector collapses to a much smaller gap late in the schedule near $D \approx 1$, and no interior crossing survives in the deep easy-plane sector. Instantaneous-state populations and the mean $m_z = 0$ occupation show that this restructuring opens stepwise reconfiguration pathways through the intermediate level and determines where finite-time diabatic losses occur. Across the disordered ensembles the quantum and classical responses to the anisotropy separate near $D \approx 1$, where the landscape develops a near-degenerate manifold of mixed $m_z = 0$ configurations: the optimized time-to-solution of the annealer falls or is unchanged relative to $D=0$ while that of both classical solvers rises, even though the exactly computed classical escape barriers fall sharply through this point rather than rising. This separation is independent of any cost-conversion convention between the protocols, and identifies the intermediate-state manifold as a structured transport channel for coherent dynamics and a flat, highly degenerate region for local classical search. Deep in the easy-plane sector the cost landscape itself becomes funnel-like and all three solvers improve, the annealer considerably more. The results establish single-ion anisotropy as a spectral control parameter of multilevel quantum annealing, and delineate the class of ternary optimization problems with an on-site localization term on which spin-1 annealing is dynamically favored.
\end{abstract}

\maketitle

\section{Introduction}
\label{intro}

Combinatorial optimization arises across physics, computer science, and engineering, with applications from materials and circuit design to logistics, portfolio optimization, and scheduling~\cite{Harrison2022max3cutcoherent,Weggemans2022solvingcorrelation}. Many such problems can be cast as minimizing a cost function over discrete variables, naturally mapping to the ground state of an effective Ising-type Hamiltonian with local fields and pairwise couplings~\cite{Finnila1994,Kadowaki1998,Farhi2001,Das2008,Lucas2014}. Since the configuration space grows exponentially with system size, classical methods can struggle on rugged, frustrated landscapes with many nearly degenerate local minima.

Quantum annealing (QA) is a heuristic approach to such problems that leverages quantum fluctuations~\cite{Kadowaki1998,Albash2018Review,Rajak2023QuantumAnnealingOverview}. The protocol initializes the system in the ground state of a simple driver Hamiltonian and then interpolates toward the problem Hamiltonian under a time-dependent schedule. In the coherent, adiabatic limit the system follows the instantaneous ground state, provided the evolution is sufficiently slow. In practice the runtime is constrained by the minimum spectral gap encountered along the annealing path~\cite{Farhi2001,Albash2018Review,Jansen2006BoundsFT}; small gaps require long schedules and increase susceptibility to nonadiabatic excitations, decoherence, and control errors~\cite{Bottarelli2025symmetrycd,Imoto2022TwistedFields,Chancellor2017LocalSearches}.

Most theoretical and experimental QA platforms use effective spin-$1/2$ degrees of freedom (qubits), which directly encode binary decision variables~\cite{Das2008,Albash2018Review,Hauke2020,Johnson2011,DenchevPRX2016,BoixoNC2016}. Many practically relevant problems are, however, intrinsically multivalued: ternary decisions (e.g.\ buy/hold/sell) and formulations with variables in sets such as $\{-1,0,+1\}$ are common. On qubit-based architectures, representing multivalued logic typically requires embedding each variable into several qubits together with penalty terms enforcing logical constraints~\cite{Weggemans2022solvingcorrelation,Mirkarimi2024LinearPenalties}. Such embeddings increase hardware overhead, reduce effective connectivity, and complicate the energy landscape, potentially shrinking relevant gaps and amplifying sensitivity to noise.

Spin-$1$ systems in particular have three eigenstates of $S^z$ labeled by $m_z=-1,0,+1$, enabling a direct representation of ternary variables without elaborate embeddings. This has motivated interest in multilevel quantum optimization and in how extra levels affect adiabatic protocols~\cite{Amin2013Qudits,WangHuSandersKais2020,Weggemans2022solvingcorrelation,Bottarelli2025symmetrycd,Mori2024Adiabatic,Perera2020,Zobov2021,Okada2019HalfHotPotts}. Beyond compact encoding, spin-$1$ models also introduce control parameters absent in spin-$1/2$ systems. A key example is the single-ion anisotropy term $D\sum_i (S_i^z)^2$, familiar from spin-$1$ magnetism such as the Blume--Capel model~\cite{Blume1966,Capel1966,Yang2008Spin1Ising}, which tunes the relative energetics of the $m_z=0$ and $m_z=\pm1$ states. In contrast to studies of variational quantum optimization or quantum approximate optimization algorithms~\cite{Weggemans2022solvingcorrelation,Bottarelli2025symmetrycd,Amin2013Qudits,WangHuSandersKais2020}, its role in quantum annealing has not been systematically analyzed, although it modifies the energetic ordering of the local states and can qualitatively reshape the avoided crossings encountered during the anneal.

Our central idea is that the intermediate $m_z = 0$ level can enable stepwise reconfiguration pathways between configurations that would otherwise require collective spin flips. When $D$ favors occupation of the $m_z=0$ state, the system may traverse configuration space via sequences of smaller transitions rather than a single large rearrangement: a site can pass from $m_z=-1$ to $m_z=+1$ through the intermediate configuration with $m_z=0$, so that one collective flip is replaced by two smaller steps, each with its own avoided crossing. This redistributes diabatic transitions over multiple crossings and thereby modifies the finite-time dynamics; whether it translates into a practical performance gain depends on the classical dynamics used as a reference, a question we address quantitatively below.

The mechanism is fundamentally different from bifurcation-based annealing approaches using nonlinear oscillators or Kerr parametric oscillators~\cite{goto2016bifurcation,goto2019kpo,puri2017kpo,nigg2017robust,goto2019simulated,kanao2021kpo}, in which candidate solutions emerge through continuous-amplitude bifurcations of nonlinear oscillator modes. Related continuous-variable approaches optimize by descending a continuously deformed landscape~\cite{Yamamura2024GeometricLandscapeAnnealing} or by annealing over continuous degrees of freedom directly~\cite{Arai2023ContinuousVariableQA}. By contrast, the present mechanism operates in a finite-dimensional spin-$1$ Hilbert space and is mediated by discrete transport through the intermediate $m_z=0$ level.

To investigate it we study anisotropic quantum annealing (AQA) in spin-$1$ Ising chains, for clean chains and for ensembles of chains with random couplings and longitudinal fields. We characterize the mechanism through spectral and dynamical diagnostics and we assess finite-time performance on the same trit cost function against two classical baselines, parallel tempering and random-restart steepest descent. Clean one-dimensional chains with finite-range interactions provide a controlled setting in which the anisotropy-controlled finite-time dynamics can be resolved and tested across disorder realizations. Throughout, statements that AQA performs better or worse than a classical protocol are strictly operational, referring to the specific dynamics, instances, and cost model defined in Sec.~\ref{Methods}.

The paper is organized as follows. Sec.~\ref{tf} introduces the spin-$1$ annealing Hamiltonian and the classical dynamics used as baselines. Sec.~\ref{Methods} describes the annealing protocol, the disorder ensembles, the classical baselines, and the performance metrics. Sec.~\ref{Results} presents the mechanism diagnostics, the ensemble statistics, and the comparison with the classical baselines. Sec.~\ref{Con} summarizes our findings and outlines directions for future work. Appendix~\ref{App:Ruggedness} analyses the classical energy landscape, and Appendix~\ref{App:Ratio} reports the convention-dependent absolute cost ratio.

\section{Theoretical Framework}\label{tf}

We consider a spin-$1$ quantum annealer, encoding each trit in the local $S^z$ eigenbasis $\{|+1\rangle,|0\rangle,|-1\rangle\}$, satisfying
\begin{align}
S^z |m_z\rangle = m_z|m_z\rangle,\quad m_z\in\{-1,0,+1\}.
\end{align}
In this basis the spin operators are
\begin{align}
S^z=
\begin{pmatrix}
1&0&0\\[2pt]
0&0&0\\[2pt]
0&0&-1
\end{pmatrix},
\quad
S^x=\frac{1}{\sqrt{2}}
\begin{pmatrix}
0&1&0\\[2pt]
1&0&1\\[2pt]
0&1&0
\end{pmatrix},
\label{eq:spin1_ops}
\end{align}
with $[S^\alpha,S^\beta]=i\epsilon_{\alpha\beta\gamma}S^\gamma$.

The system evolves under the standard interpolating annealing Hamiltonian
\begin{align}
H(t) = s(t) H_P + (1-s(t))\,H_D,
\quad
s(t)=\frac{t}{t_f}\in[0,1],
\label{eq:anneal_H}
\end{align}
where $t_f$ is the total annealing time, the problem Hamiltonian is
\begin{align}
H_P &= -\sum_{\langle i,j\rangle} J_{ij}\, S_{i}^{z}S_{j}^{z}
      - \sum_{i} h_i\, S_{i}^{z}
      + D\sum_{i}(S_{i}^{z})^{2},
\label{eq:HP_def}
\end{align}
and the time-independent driver Hamiltonian is
\begin{align}
H_D &= -\sum_{i}S_{i}^{x}.
\label{eq:HD_def}
\end{align}
Here $S_{i}^{\alpha}$ are spin-$1$ operators acting on site $i$; $J_{ij}$ are nearest-neighbor couplings, uniform ($J_{ij}=J$) for clean chains and random for the disordered ensembles; $h_i$ are longitudinal fields, likewise uniform or random; and $D$ is the single-ion anisotropy parameter. Both endpoints of the schedule are pinned: at $t=0$ the Hamiltonian is $H_D$, whose ground state is a product of single-site $S^x$ eigenstates and serves as the exact initial state, while at $t=t_f$ it is exactly $H_P$.

By construction $H_P$ is the quantum embedding of the classical cost function: on computational basis states $|s_1,\dots,s_N\rangle$ with $S_i^z|s_i\rangle = s_i|s_i\rangle$ and $s_i\in\{-1,0,+1\}$,
\begin{align}
H_P |s_1,\dots,s_N\rangle = H(s) |s_1,\dots,s_N\rangle .
\end{align}

Let $\{\ket{\psi_k(t)},E_k(t)\}$ be the instantaneous spectrum of $H(t)$ with ground state $k{=}0$. The adiabatic bound~\cite{Jansen2006BoundsFT} gives
\begin{align}
1-P_{gs} \lesssim \max_t
\frac{\big|\langle\psi_1(t)|\dot H(t)|\psi_0(t)\big\rangle\big|^2}{\Delta(t)^4},
\label{eq:adiabatic}
\end{align}
with gap $\Delta(t)=E_1(t)-E_0(t)$. For the linear schedule of Eq.~\eqref{eq:anneal_H}, $\dot H = (H_P - H_D)/t_f$, so diabatic effects are suppressed as $1/t_f^2$ away from small gaps.
Near an avoided crossing with minimum gap $\Delta_{\min}$, a two-level linearization yields the Landau--Zener formula~\cite{Landau}
\begin{align}
P_{\rm exc}\approx \exp\left(-\frac{\pi\Delta_{\min}^2}{2|\alpha|}\right),
\label{eq:LZ}
\end{align}
where $\alpha$ is the diabatic slope difference ($\hbar=1$). By an avoided crossing we mean an interior local minimum of the instantaneous gap: a point $0<s^{*}<1$ at which $\Delta(s)$ closes to a minimum $\Delta_{\min}$ and reopens, signaling a near-resonance of two instantaneous eigenstates whose characters exchange across $s^{*}$. The anisotropy term can modify both the location and the structure of these crossings by introducing intermediate low-energy states associated with the $m_z=0$ level.

We use the term trit for a classical three-valued variable $s\in\{-1,0,+1\}$, so that the classical configuration space is $\mathcal{S}=\{-1,0,+1\}^N$ with cost function
\begin{align}
H(s) &= -\sum_{\langle i,j\rangle} J_{ij}\, s_i s_j
        - \sum_{i=1}^N h_i\, s_i
        + D\sum_{i=1}^N s_i^2 .
\label{eq:classical_cost}
\end{align}

The anisotropy enters Eq.~\eqref{eq:classical_cost} on the same footing as the couplings and the fields: $D\sum_i s_i^2$ is an on-site localization term of the ternary cost function, penalizing ($D>0$) or rewarding ($D<0$) occupation of the intermediate value $s_i=0$. The $D$ term is is an intrinsic part of the ternary cost function. Varying $D$ therefore generates a parameterized family of optimization landscapes, each with its own ground state and low-energy structure. All comparisons below are made at fixed $D$, between protocols acting on the identical cost function and the identical disorder realizations; the anisotropy dependence of each protocol is then assessed relative to that protocol's own performance at $D=0$.

We adopt the standard terminology of single-ion anisotropy in magnetism: $D<0$ defines the axial (easy-axis) sector, in which the $m_z=\pm1$ states are locally favored and $m_z=0$ is penalized; $D>0$ defines the easy-plane sector, in which $m_z=0$ is favored. The crossover region $D\approx1$, where local $m_z=0$ occupation becomes energetically competitive with $m_z=\pm1$ without either sector dominating, plays a central role in what follows.

One exact consequence of this structure is worth stating. On the subspace spanned by configurations with $m_z=\pm1$ on every site, the anisotropy term takes the constant value $DN$ and contributes only an overall energy shift, so a dynamics confined to that subspace would be strictly independent of $D$. Every anisotropy dependence reported in this work is therefore mediated by amplitude in the intermediate $m_z=0$ level, on both the quantum and the classical side.

The elementary move used in classical baselines is the single-site Metropolis update~\cite{Metropolis1953,Kirkpatrick1983} at inverse temperature $\beta=1/T$:
\begin{enumerate}
  \item Pick a site $i\in\{1,\dots,N\}$ uniformly at random.
  \item Propose a new local value $s_i'\in\{-1,0,+1\}\setminus\{s_i\}$ uniformly over the two
        alternatives (the proposal distribution is symmetric).
  \item Let $s'$ be $s$ with $s_i$ replaced by $s_i'$, and define the local energy for setting
        $s_i=m$ by
  \begin{align}
    \varepsilon_i(m) = -m \sum_{j\in\partial i} J_{ij}\, s_j - h_i m + Dm^2,
  \end{align}
  where $\partial i$ denotes the neighbors of $i$. The energy change is
  \begin{align}
    \Delta E &= H(s')-H(s) = \varepsilon_i(s_i') - \varepsilon_i(s_i)\\
    &= -\big(s_i'-s_i\big)\!\left(\sum_{j\in\partial i} J_{ij}\, s_j + h_i\right)
       + D\big[(s_i')^2 - s_i^2\big].
  \end{align}
  \item Accept with Metropolis probability
        $P_{\rm acc}(s \to s') = \min\{1, e^{-\beta \Delta E}\}$; otherwise keep $s$.
\end{enumerate}

The primary classical baseline is replica-exchange Monte Carlo, or parallel tempering (PT)~\cite{SwendsenWang1986,HukushimaNemoto1996}, which simulates $R$ replicas $\{s^{(1)},\dots,s^{(R)}\}$ at fixed inverse temperatures $\beta_1 > \dots > \beta_R$, targeting
\begin{align}
\Pi\bigl(s^{(1)},\dots,s^{(R)}\bigr) \propto
\prod_{r=1}^{R} e^{-\beta_r H(s^{(r)})}.
\label{eq:PT_ensemble}
\end{align}
Each PT sweep consists of one Metropolis sweep of every replica at its own temperature, followed by
exchange attempts between replicas at neighboring temperatures, the configurations of replicas $r$
and $r+1$ being swapped with probability
\begin{align}
P_{\rm swap}(r \leftrightarrow r+1)
= \min\Bigl\{1, e^{(\beta_r-\beta_{r+1})\left[E_r - E_{r+1}\right]}\Bigr\},
\label{eq:PT_swap}
\end{align}
where $E_r = H(s^{(r)})$. This rule satisfies detailed balance with respect to Eq.~\eqref{eq:PT_ensemble}, so each replica samples its own Boltzmann distribution while configurations random-walk in temperature: hot replicas cross barriers that would trap a single cold chain, and cold replicas refine candidate minima. For optimization purposes a PT run counts as successful when the lowest energy encountered in any replica reaches the exact ground-state energy within a fixed numerical tolerance. The replica count, temperature ladder, sweep budgets, acceptance diagnostics, second baseline, and cost accounting are specified in Sec.~\ref{Methods}.

\section{Methods}\label{Methods}

\subsection{Systems and instances}

We study one-dimensional nearest-neighbor open chains of length $N\in\{6,7,10\}$, each site carrying a spin-$1$ degree of freedom, so that $\dim\mathcal{H}=3^{N}$ remains small enough for numerically exact treatment of the full quantum dynamics.

Clean chains, used for the mechanism diagnostics, have uniform couplings $J_{ij}=J$ and a small uniform field $h_i=h=0.2$. All clean-chain diagnostics use the ferromagnetic sign $J=+1$, for which the weak symmetry-breaking field selects a unique classical ground state at every $D$. The antiferromagnetic clean chain is unsuitable for the same purpose at even $N$: the two N\'eel configurations both carry zero total magnetization, so a uniform field does not lift their degeneracy and the instantaneous gap closes at the end of the schedule. Disordered ensembles, used for the performance statistics, consist of $M=200$ instances per chain length with independent random couplings $J_{ij}\sim\mathcal{N}(0,1)$ and fields $h_i\sim\mathcal{N}(0,0.5^{2})$.

All protocols are evaluated on the same disorder realizations, generated from identical pseudorandom seeds, so every ensemble comparison below is paired. For each instance we verify that the diagonal matrix elements of $H_P$ reproduce the classical cost function of Eq.~\eqref{eq:classical_cost}; the largest discrepancy over all $6600$ (instance, $D$) pairs is $2\times10^{-14}$. The anisotropy is scanned over $D\in[-5,5]$ on an integer grid for both families. Exact ground-state energies are obtained by dynamic programming over the chain (transfer-matrix minimization), which is exact for finite-range one-dimensional problems at any $N$, and provide the ground truth for all success criteria. The same recursion returns the degeneracy: with continuous random couplings and fields exact ties occur with probability zero, and indeed every instance at every anisotropy is found to have a one-dimensional ground space, in agreement with the degeneracy obtained independently from the diagonal of $H_P$. No $D$-dependent change of degeneracy can therefore contribute to the anisotropy dependence reported below.

\subsection{Quantum protocol and diagnostics}

For AQA we propagate the time-dependent Schr\"odinger equation for Eq.~\eqref{eq:anneal_H} by a second-order split-operator method ~\cite{Childs2021Trotter}. The schedule is discretized into $n_{\rm step}$ equal intervals of length $\delta t = t_f/n_{\rm step}$, with $s$ evaluated at the midpoint of each interval, giving the update
\begin{align}
U_k = e^{-\tfrac{i}{2}\delta t\, s_k H_P}\,
      e^{-i \delta t (1-s_k) H_D}\,
      e^{-\tfrac{i}{2}\delta t\, s_k H_P},
\label{eq:strang}
\end{align}
with $s_k = (k+\tfrac12)/n_{\rm step}$. Since $H_P$ is diagonal in the computational basis and $H_D$ is a sum of single-site terms, both factors are applied exactly and the propagation is unitary by construction; norm conservation therefore carries no information about the discretization error, and convergence is established by step doubling instead. With $n_{\rm step}=160$, used for the whole time-to-solution grid, recomputing each instance at $2n_{\rm step}$ changes the success probability at the optimal annealing time by $3\times10^{-3}$ on average ($99$th percentile $4\times10^{-2}$), and changes $TTS^{opt}$ by a median of $8\times10^{-4}$ ($90$th percentile $4\times10^{-2}$); the largest drifts occur in the deep axial sector, where the anneal is least adiabatic. The near-adiabatic diagnostic of Sec.~\ref{Results}~B is a separate calculation at $t_f=10^{3}$ and needs a much finer discretization at fixed $t_f$; it uses $n_{\rm step}=8000$, and its $p_0(s)$ trace changes by at most $9\times10^{-4}$ when $n_{\rm step}$ is halved.

Success is measured against the full ground space of $H_P$: with $\mathcal{P}_0$ the projector onto all computational basis states whose classical cost lies within $10^{-9}$ of the exact ground-state energy,
\begin{align}
P_{AQA}(t_f) = \bra{\psi(t_f)} \mathcal{P}_0 \ket{\psi(t_f)} .
\label{eq:P_AQA}
\end{align}
Using a projector rather than a single-state fidelity keeps the criterion well defined irrespective of degeneracy.

For the clean-chain diagnostics we record, along the anneal: (i) the instantaneous gap $\Delta(t)=E_1(t)-E_0(t)$, on a uniform grid of $161$ schedule points (spacing $6.25\times10^{-3}$), by sparse diagonalization of $H(t)$ shifted so that the two lowest levels become the two largest in magnitude; (ii) the population of the instantaneous ground state $|\langle\psi_0(t)|\psi(t)\rangle|^{2}$; and (iii) the mean occupation of the intermediate level,
\begin{align}
p_0(t) = \frac{1}{N}\sum_{i=1}^{N}
\bra{\psi(t)}\, \bigl(\ket{0}\!\bra{0}\bigr)_i\, \ket{\psi(t)},
\label{eq:pop0}
\end{align}
which directly witnesses transport through the $m_z=0$ level. Every interior local minimum of $\Delta(s)$ found on the coarse grid is refined to an interval of $10^{-6}$ in $s$; all minimum gaps quoted below are refined values, and the coarse-grid spacing is stated so that the improvement over the sampled values can be judged.

\subsection{Classical baselines}

\subsubsection{Parallel tempering}

The primary baseline is the PT algorithm of Sec.~\ref{tf}, applied to the identical trit cost function. Rather than fixing the replica count we treat it as a tuning parameter and scan $R\in\{2,4,8,16\}$, rebuilding for each $R$ a geometric temperature ladder between $T_{\min}=0.05$ and $T_{\max}=3.0$. Replicas are initialized uniformly at random from the $3^{N}$ classical states, and exchange attempts alternate between even and odd neighboring pairs. A run counts as successful when the lowest energy encountered in any replica reaches the exact ground-state energy within the same $10^{-9}$ tolerance used on the quantum side.

For each instance, anisotropy and replica count, PT is run at $14$ logarithmically spaced sweep budgets between $1$ and $10^{3}$, with $64$ independent repetitions per budget from which the success probability is estimated. The work of a budget of $S$ sweeps is $W = S N R$ elementary single-site updates, and the full budget is charged even when a run reaches the ground state early, matching the fixed-budget restart convention behind Eq.~\eqref{eq:tts}. The reported $TTS^{opt}_{PT}$ is minimized over both the sweep budget and the replica count, so the baseline is tuned rather than fixed.

Charging every replica makes the tuned optimum favor short ladders: $R=2$ minimises the time-to-solution for $34$--$35\%$ of (instance, $D$) pairs, $R=4$ for $30$--$35\%$, $R=8$ for $19$--$23\%$ and $R=16$ for only $11$--$13\%$, drifting weakly toward larger $R$ as $N$ increases. The exchange acceptance is correspondingly a strong function of $R$, lying in $[0.78,0.91]$ at $R=16$ but only $[0.00,0.15]$ at $R=2$. For $R=2$, the optimal configuration behaves approximately as two weakly coupled chains rather than as a well-mixed replica-exchange ladder. Nevertheless, it minimizes the time-to-solution under the stated work accounting and is therefore retained in the optimized classical baseline. It is, however, the configuration that minimizes the time-to-solution under the stated cost model, which is the relevant baseline for an optimization comparison, and reporting the full $R$ scan makes the trade-off between mixing quality and charged work explicit.

Two finite-sample effects enter the optimization of the classical baseline. Selecting the sweep budget and replica count from finitely many repetitions produces a downward bias in the estimated optimum. Cross-validation increases the median TTS by $2$--$25\%$ while leaving the conclusions of Sec.~\ref{Results} unchanged. We therefore retain the estimator obtained by direct minimization over the tested budgets and replica counts in the figures. The second effect concerns the position of the optimum on the two grids. All AQA optima are interior to the $t_f$ grid, falling between its third and twelfth values. On the classical side the optimum falls on the smallest budget of one sweep for $1.14\%$ of (instance, $D$) pairs ($75$ of $6600$), all in the axial sector $D\le0$ and distributed as $54$, $16$ and $5$ across $N=6,7,10$. Since the grid cannot extend below a single sweep under sweep-granular accounting, these are upper bounds on the optimum a finer granularity would give; the affected instances are among the easiest of their ensembles, and excluding them shifts the PT median at the affected $(N,D)$ by at most $2.5\%$.

\subsubsection{Random-restart steepest descent}

A single classical algorithm cannot distinguish a property of the cost landscape from a property of thermal sampling, so we add a second, structurally different baseline: random-restart steepest descent (SD), a greedy rather than thermal dynamics. As described in Appendix~\ref{App:Ruggedness}, exact enumeration yields for every instance the ground-basin fraction $f_{gs}(N,D)$ of Eq.~\eqref{eq:fgs}, the fraction of configurations whose steepest-descent path terminates in the ground space, together with the mean number $w_{\rm desc}$ of single-site energy evaluations consumed by one descent. A restart from a uniformly random configuration therefore succeeds with probability $f_{gs}$ at cost $w_{\rm desc}$, and
\begin{align}
TTS_{SD}(N,D) = w_{\rm desc}\,
\frac{\ln(1-p_{target})}{\ln\bigl(1-f_{gs}(N,D)\bigr)},
\label{eq:tts_greedy}
\end{align}
with $w_{\rm desc}$ in the same elementary single-site units used for PT. In practice $w_{\rm desc}$ ranges over $34$--$39$ at $N=6$, $43$--$49$ at $N=7$ and $74$--$81$ at $N=10$ across the anisotropy range. Eq.~\eqref{eq:tts_greedy} is exact for this ensemble rather than estimated, and carries no selection bias.

\subsection{Performance metrics and cost accounting}

All three solvers are evaluated through the time-to-solution (TTS)~\cite{ronnow2014},
\begin{align}
TTS(\tau) = \tau\,
\frac{\ln(1-p_{target})}{\ln\bigl(1-P(\tau)\bigr)},
\qquad p_{target} = 0.95,
\label{eq:tts}
\end{align}
where $\tau$ is the cost of a single run in the protocol's native units and $P(\tau)$ the corresponding success probability; when $P(\tau)\ge p_{target}$ a single run suffices and $TTS=\tau$. For AQA, $\tau$ is the annealing time $t_f$, optimized over $14$ logarithmically spaced values in $[1,10^{3}]$; for PT, $\tau$ is the work $W$, optimized over the sweep-budget grid and the replica count; for SD, $\tau=w_{\rm desc}$ and no optimization is required. The optimized value $TTS^{opt}=\min_\tau TTS(\tau)$ is recorded per instance, protocol and anisotropy.

The two success criteria are not identical, and the difference does not act in the same direction on the two sides. The quantum criterion is single-shot: $P_{AQA}$ is the probability that a measurement at $t_f$ returns a ground state, and any earlier passage through the ground space is irrelevant. A PT run, by contrast, succeeds if the ground energy is encountered in any replica at any point, while the full budget is charged either way, so the PT criterion is the more permissive of the two and again favours the classical baseline. Rescoring PT on the single configuration held by its coldest replica at the end of the budget (the direct analogue of the quantum readout) raises the medians, as expected, but leaves the anisotropy dependence intact: at $D=1$ the rescored ratio to $D=0$ is $1.00$, $1.20$ and $1.15$ at $N=6,7,10$, against $1.02$, $1.24$ and $1.12$ under the permissive criterion.

Because the native cost units of the protocols are not directly commensurable, we report results in three complementary forms. First, per-protocol TTS statistics as functions of $D$ and $N$ in each protocol's own units (Fig.~\ref{fig:tts_paired}); all statements about the anisotropy dependence of performance rest on these. Second, the cross-protocol comparison through the doubly normalized ratio
\begin{align}
\tilde r(N,D) =
\frac{TTS^{opt}_{C}(N,D)\big/TTS^{opt}_{C}(N,0)}
     {TTS^{opt}_{AQA}(N,D)\big/TTS^{opt}_{AQA}(N,0)},
\label{eq:rtilde}
\end{align}
where $C$ denotes either classical solver. Any constant $\kappa$ relating one unit of annealing time to one elementary spin update cancels identically in Eq.~\eqref{eq:rtilde}, so $\tilde r$ is free of the cost-conversion convention; it is the only cross-protocol quantity quoted in the main text. Third, the convention-dependent absolute ratio $r(N,D)=\kappa\,TTS^{opt}_{PT}/TTS^{opt}_{AQA}$ is reported in Appendix~\ref{App:Ratio}, labeled there as depending on a declared $\kappa$.

For each $(N,D)$ we report the median and interquartile range of $TTS^{opt}$ over all $M$ instances, together with the solved-instance fraction. Unsolved instances are assigned infinite TTS and retained in the order statistics, so the median is finite only when more than half the ensemble is solved; this avoids the optimistic bias of conditioning on successful instances. In the ensembles studied the solved fraction is unity at every $(N,D)$ for all three solvers, so the convention never binds. Because the solvers are evaluated on identical realizations, the anisotropy dependence of each is quantified by the median of the per-instance ratio $TTS^{opt}(D)/TTS^{opt}(D{=}0)$ rather than by the ratio of medians, with significance assessed by a Wilcoxon signed-rank test against the paired $D=0$ values. All medians, median ratios and values of $\tilde r$ carry $95\%$ percentile bootstrap confidence intervals over the $M$ instances ($10^4$ resamples; for $\tilde r$ numerator and denominator are resampled jointly, preserving the pairing).

\section{Results}\label{Results}

\subsection{Spectral restructuring}

\begin{figure}[tb]
\includegraphics[width=\columnwidth]{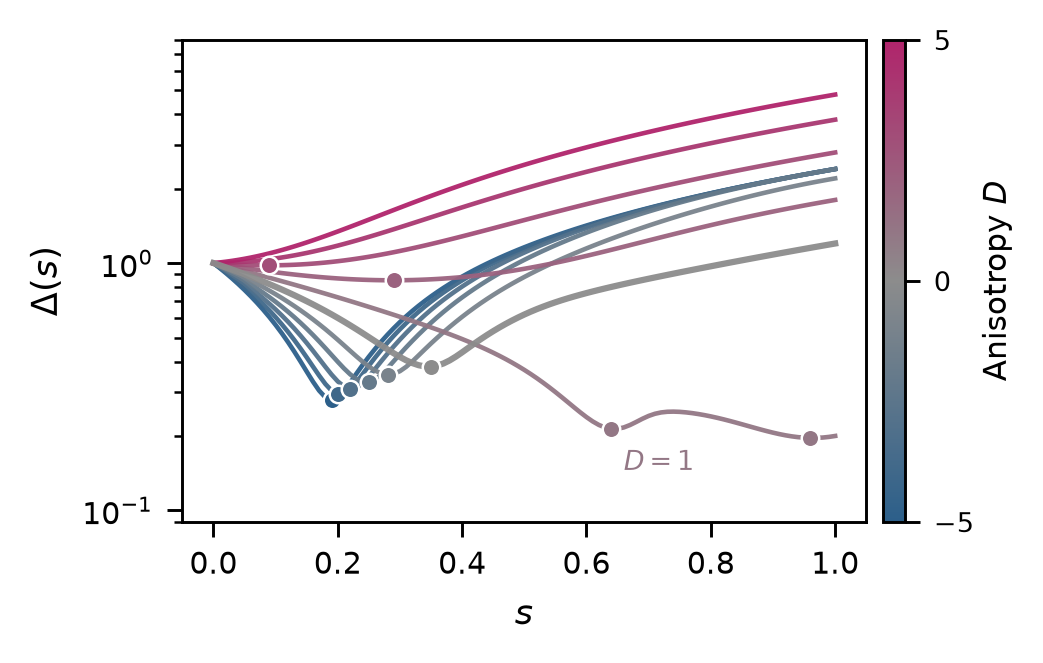}
\caption{Instantaneous gap $\Delta(s)$ of the clean chain at $N=10$ ($J=1$, $h=0.2$) over the full anisotropy sweep $D=-5,\dots,5$ (color scale). Circles mark the interior local minima of each trace, i.e.\ the avoided crossings traversed by the anneal. The highlighted $D=1$ trace exhibits two distinct avoided crossings, at $s^{*}=0.64$ and $s^{*}=0.96$; for $D\ge4$ the gap is monotonic in $s$ and no crossing occurs.}
\label{fig:gaptrace}
\end{figure}

\begin{figure}[tb]
\includegraphics[width=\columnwidth]{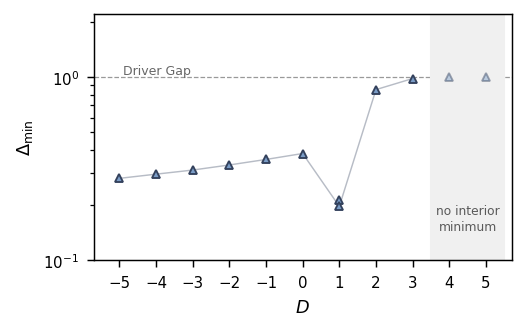}
\caption{Refined minimum gap $\Delta_{\min}$ versus anisotropy at $N=10$. Each symbol is an interior local minimum of the corresponding trace in Fig.~\ref{fig:gaptrace}; the two symbols at $D=1$ are the two avoided crossings of the split transition. For $D\ge4$ (shaded) no interior minimum exists: the smallest gap along the path is the driver-endpoint value, which equals $1$.}
\label{fig:gapmin}
\end{figure}

We first characterize the mechanism on clean chains, where the spectral structure can be resolved completely. Figure~\ref{fig:gaptrace} shows the instantaneous gap along the anneal for the full anisotropy sweep, and Fig.~\ref{fig:gapmin} condenses each trace into its minimum-gap structure. Avoided crossings are identified as interior local minima of $\Delta(s)$, in the operational sense of Sec.~\ref{tf}. Where $\Delta(s)$ is monotonic the anneal encounters no crossing, and the smallest gap along the path is the driver-endpoint value $\Delta(0)$, which equals unity independently of $N$: the spectrum of a single-site $-S^{x}$ is $\{-1,0,+1\}$, so the cheapest excitation of $H_D$ raises one site from $-1$ to $0$ at unit cost. Values quoted below are for $N=10$, the size shown in the figures, unless stated otherwise; the other two sizes agree to within a few percent except where noted.

Three regimes emerge along the anisotropy axis. In the axial sector, $D\le0$, every trace has a single mid-schedule crossing whose depth depends only weakly on $D$, rising from $\Delta_{\min}=0.28$ at $D=-5$ to $0.38$ at $D=0$ while drifting to later schedule times, $s^{*}=0.19\to0.35$. The rigidity is expected: the relevant low-energy states are composed of $m_z=\pm1$ configurations, on which the single-ion term acts as the constant $DN$, so varying $D$ shifts the spectrum without reorganizing it.

Near degeneracy the picture changes qualitatively. At $D=1$ the mid-schedule crossing is replaced by structure late in the schedule and at roughly half the gap, and here the three sizes differ. At $N=10$ there are two interior minima, at $s^{*}=0.64$ and $0.96$ with $\Delta_{\min}=0.21$ and $0.20$; at $N=7$ likewise two, at $s^{*}=0.70$ and $0.96$ with $\Delta_{\min}=0.17$ and $0.20$. At $N=6$ the same collapse occurs but resolves into a single, deeper late-schedule minimum, at $s^{*}=0.75$ with $\Delta_{\min}=0.12$. The resolution into two distinct crossings is thus a feature of the two larger sizes and not of the smallest, whereas the collapse of the gap and its migration to late schedule times is common to all three; we report this size dependence rather than extrapolating from it. The late crossing at $s^{*}=0.96$ is a genuine interior minimum and not an endpoint artifact: it lies six coarse-grid points from the end of the schedule, with $\Delta(s)$ increasing on both sides. All gaps quoted here are golden-section refinements of minima located on a grid of spacing $6.25\times10^{-3}$, and the refinement shifts them by less than $10^{-3}$ throughout.

In the easy-plane sector the crossing first reopens, with $\Delta_{\min}=0.85$ at $D=2$ and $0.98$ at $D=3$, and then disappears altogether: for $D\ge4$ the gap grows monotonically from the driver value and the anneal traverses no crossing at all. The interior minimum at $D=3$ therefore lies only marginally below the driver gap and has migrated to $s^{*}=0.087$, near the start of the schedule; it is a residual feature rather than a bottleneck.

\subsection{Intermediate-state transport and diabatic leakage}

\begin{figure}[tb]
\includegraphics[width=\columnwidth]{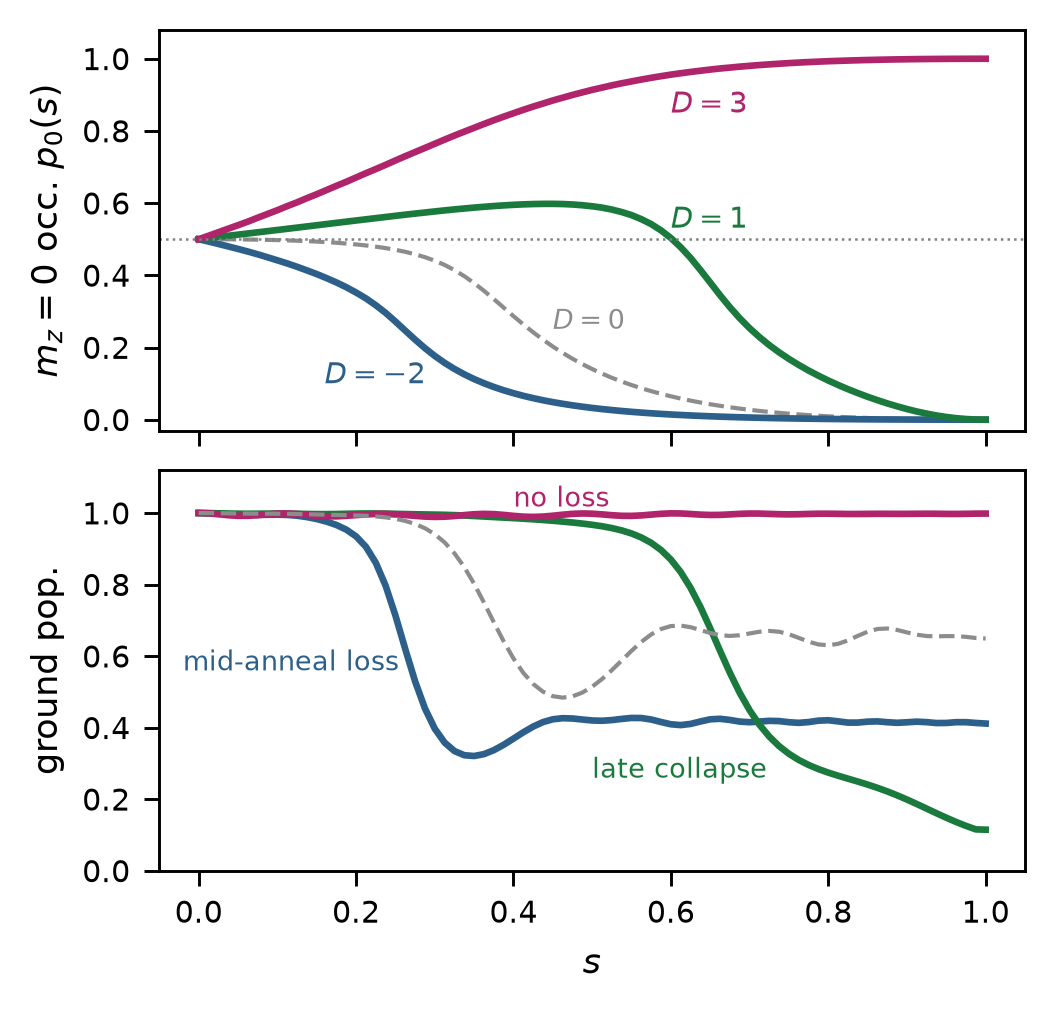}
\caption{Transport through the intermediate level and its finite-time cost, on the clean chain at
$N=10$, for the three regime representatives, axial ($D=-2$), near-degenerate ($D=1$), easy-plane
($D=3$), and the $D=0$ reference (dashed). Top: mean $m_z=0$ occupation $p_0(s)$,
Eq.~\eqref{eq:pop0}, along a near-adiabatic anneal ($t_f=10^{3}$, $n_{\rm step}=8000$); all
trajectories start at the driver value $p_0=1/2$ (dotted). Bottom: population of the instantaneous
ground state along a fast anneal ($t_f=30$). The panels share the schedule axis $s=t/t_f$, so
features at the same $s$ are vertically aligned but correspond to different annealing times.}
\label{fig:mech}
\end{figure}

The dynamical counterpart of this restructuring is shown in Fig.~\ref{fig:mech}. Along a near-adiabatic anneal the mean $m_z=0$ occupation interpolates from its driver value $p_0=1/2$ to the ground-state value of $H_P$: in the axial case it decays monotonically to zero, at $D=0$ it commits by mid-schedule, and at $D=3$ it grows to unity as the all-zeros configuration is approached. The mechanistically significant case is $D=1$, where $p_0$ instead rises above its driver value, peaks at $0.60$ near mid-schedule, and stays elevated until it falls back through $1/2$ at $s\approx0.6$. The system thus occupies the intermediate level throughout the schedule region in which the classical spectrum is nearly degenerate (Appendix~\ref{App:Ruggedness}), and releases it as it traverses the first of the two avoided crossings of Fig.~\ref{fig:gaptrace}: a direct dynamical witness of intermediate-state transport, independent of the spectral evidence.

The same structure controls where diabatic losses occur at finite speed. At $t_f=30$, deliberately fast, the ground-state population behaves in the three regimes exactly as the crossing structure dictates. At $D=3$ it is conserved along the entire schedule, never falling below $0.99$: the interior minimum barely departs from the driver value, so the anneal meets no near-degeneracy and stays adiabatic even at this speed. At $D=-2$ the population drops once, sharply, at $s\approx0.3$, the location of the single mid-schedule crossing, and is flat elsewhere. At $D=1$ it instead survives deep into the schedule and collapses late, with the loss localized at the two small late-schedule crossings; the $D=0$ reference is intermediate between the two. Within the temporal resolution of these simulations the dominant diabatic losses are localized near the anisotropy-dependent crossings, their magnitude tracks the corresponding gap structure, and no appreciable loss is resolved away from the crossings on the schedule grid used.

The value $t_f=30$ is chosen fast enough that the diabatic losses are large enough to localize in $s$, and is used here only to resolve where along the schedule they occur. Performance comparisons below instead use the optimized time-to-solution.

\subsection{Ensemble performance and operational comparison}

\begin{figure}[tb]
\includegraphics[width=\columnwidth]{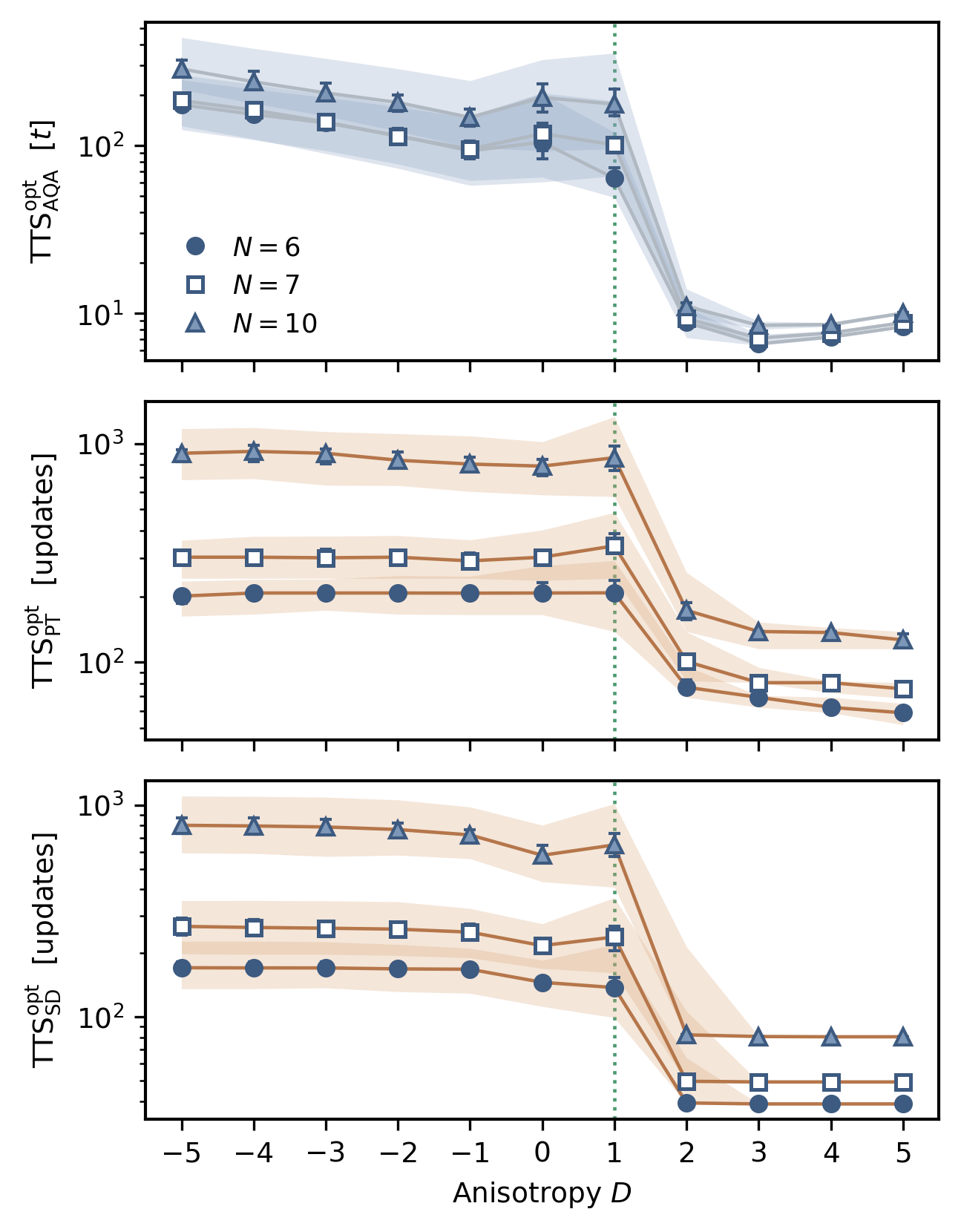}
\caption{Optimized time-to-solution over the disordered ensembles versus anisotropy $D$, for each
solver in its own native cost units: annealing time $t_f$ for AQA (top) and elementary single-site
updates for parallel tempering (middle) and random-restart steepest descent (bottom). Symbols are
medians over the $M=200$ instances per $(N,D)$; error bars are $95\%$ percentile bootstrap intervals on the median and shaded bands span the interquartile range. Dotted line: $D=1$.}
\label{fig:tts_paired}
\end{figure}

We now turn to the disordered ensembles, which test whether the clean-chain mechanism survives random couplings and fields. Every instance at every anisotropy was solved by all three solvers within the respective budget grids. Figure~\ref{fig:tts_paired} shows the optimized time-to-solution of each solver in its own cost units, and Table~\ref{tab:ratios} collects the corresponding anisotropy responses,
\begin{align}
\rho(N,D) = \mathrm{med}\left[\frac{TTS^{opt}(N,D)}{TTS^{opt}(N,0)}\right],
\label{eq:rho}
\end{align}
the median over instances of the per-instance ratio to $D=0$, together with the convention-free ratio $\tilde r$ of Eq.~\eqref{eq:rtilde}. Confidence intervals appear as error bars in the figures and are quoted in the text only where the argument turns on them.

\begin{table*}[!t]
\caption{Anisotropy response $\rho(N,D)$ of Eq.~\eqref{eq:rho} for the three solvers, and the
convention-free ratio $\tilde r(N,D)=\rho_{PT}/\rho_{AQA}$ of Eq.~\eqref{eq:rtilde}, over the
disordered ensembles. Values below unity indicate faster solution than at $D=0$ for the same
solver. All entries are medians of per-instance ratios; $95\%$ percentile bootstrap intervals are
shown as error bars in Figs.~\ref{fig:tts_paired} and~\ref{fig:rtilde}. The $D=0$ row is unity by
construction.}
\label{tab:ratios}
\begin{ruledtabular}
\begin{tabular}{ccccccccccccc}
& \multicolumn{3}{c}{$\rho_{AQA}$} & \multicolumn{3}{c}{$\rho_{PT}$}
& \multicolumn{3}{c}{$\rho_{SD}$} & \multicolumn{3}{c}{$\tilde r$ (PT)}\\
\cline{2-4}\cline{5-7}\cline{8-10}\cline{11-13}
$D$ & $N{=}6$ & $N{=}7$ & $N{=}10$ & $N{=}6$ & $N{=}7$ & $N{=}10$
    & $N{=}6$ & $N{=}7$ & $N{=}10$ & $N{=}6$ & $N{=}7$ & $N{=}10$\\
\hline
$-5$ & 1.64 & 1.75 & 1.76 & 0.91 & 1.00 & 1.13 & 1.18 & 1.22 & 1.35 & 0.55 & 0.57 & 0.64 \\
$-4$ & 1.45 & 1.53 & 1.56 & 0.93 & 1.00 & 1.16 & 1.17 & 1.21 & 1.34 & 0.64 & 0.65 & 0.74 \\
$-3$ & 1.29 & 1.34 & 1.33 & 0.95 & 1.00 & 1.12 & 1.17 & 1.18 & 1.32 & 0.73 & 0.75 & 0.84 \\
$-2$ & 1.12 & 1.14 & 1.11 & 0.97 & 1.00 & 1.09 & 1.15 & 1.18 & 1.31 & 0.86 & 0.88 & 0.98 \\
$-1$ & 0.96 & 0.96 & 0.93 & 0.94 & 0.95 & 1.03 & 1.12 & 1.14 & 1.22 & 0.98 & 0.99 & 1.11 \\
$0$  & 1.00 & 1.00 & 1.00 & 1.00 & 1.00 & 1.00 & 1.00 & 1.00 & 1.00 & 1.00 & 1.00 & 1.00 \\
$+1$ & 0.73 & 0.95 & 1.04 & 1.02 & 1.24 & 1.12 & 1.02 & 1.20 & 1.18 & 1.39 & 1.31 & 1.08 \\
$+2$ & 0.094 & 0.085 & 0.063 & 0.40 & 0.37 & 0.25 & 0.30 & 0.29 & 0.20 & 4.22 & 4.37 & 3.92 \\
$+3$ & 0.069 & 0.063 & 0.044 & 0.33 & 0.27 & 0.18 & 0.27 & 0.23 & 0.14 & 4.70 & 4.35 & 3.98 \\
$+4$ & 0.072 & 0.066 & 0.043 & 0.29 & 0.25 & 0.16 & 0.27 & 0.23 & 0.14 & 4.04 & 3.81 & 3.72 \\
$+5$ & 0.082 & 0.074 & 0.051 & 0.27 & 0.24 & 0.17 & 0.27 & 0.23 & 0.14 & 3.31 & 3.30 & 3.27 \\
\end{tabular}
\end{ruledtabular}
\end{table*}

The three solvers respond to the anisotropy in qualitatively different ways. The annealer's response is large and non-monotonic, and mirrors the crossing structure of Fig.~\ref{fig:gapmin}: $\rho_{AQA}$ grows to $1.6$--$1.8$ in the deep axial sector, returns to unity near $D=-1$, and then collapses in the easy-plane sector to $0.04$--$0.07$ at $D=3$, an improvement by a factor of $14$--$23$ over the same protocol at $D=0$ that deepens with system size. Both classical solvers respond far more weakly and with a different shape: their axial response is flat to within $\pm35\%$, and their easy-plane improvement saturates around a factor of three to seven. At these sizes steepest descent is not a weak baseline; in raw work units its median time-to-solution is comparable to that of tuned parallel tempering, and at $N=10$ and $D=0$ somewhat below it.

The easy-plane improvement reflects both the annealing dynamics and the simplification of the underlying cost landscape. As Appendix~\ref{App:Ruggedness} shows, for $D\ge3$ the classical landscape trivializes: the mean number of strict local minima drops to essentially one and the ground-basin fraction under greedy descent becomes indistinguishable from unity. Part of the improved finite-time behavior of all three solvers in this sector therefore reflects a property of that member of the problem family rather than of the dynamics acting on it.

The clearest separation between the quantum and classical responses occurs near $D \approx 1$ with a size dependent pattern. There the annealer speeds up significantly at $N=6$ ($\rho_{AQA}=0.73$, Wilcoxon $p=3\times10^{-5}$) and shows no resolvable change at $N=7$ and $N=10$ ($p=0.5$ and $0.7$), while both classical solvers slow down: parallel tempering significantly at the two larger sizes ($p=2\times10^{-5}$ and $6\times10^{-4}$), steepest descent at all three. The statement supported at every size is therefore not that the annealer accelerates while the classical solvers slow, but that the two responses separate: at $N=6$ the annealer gains while the classical solvers are flat, and at $N=7$ and $N=10$ the classical solvers lose while the annealer holds. This is the region where, as Appendix~\ref{App:Ruggedness} shows, the landscape develops a near-degenerate manifold of mixed
$m_z=0$ configurations while its escape barriers are at their weakest.

\begin{figure}[tb]
\includegraphics[width=\columnwidth]{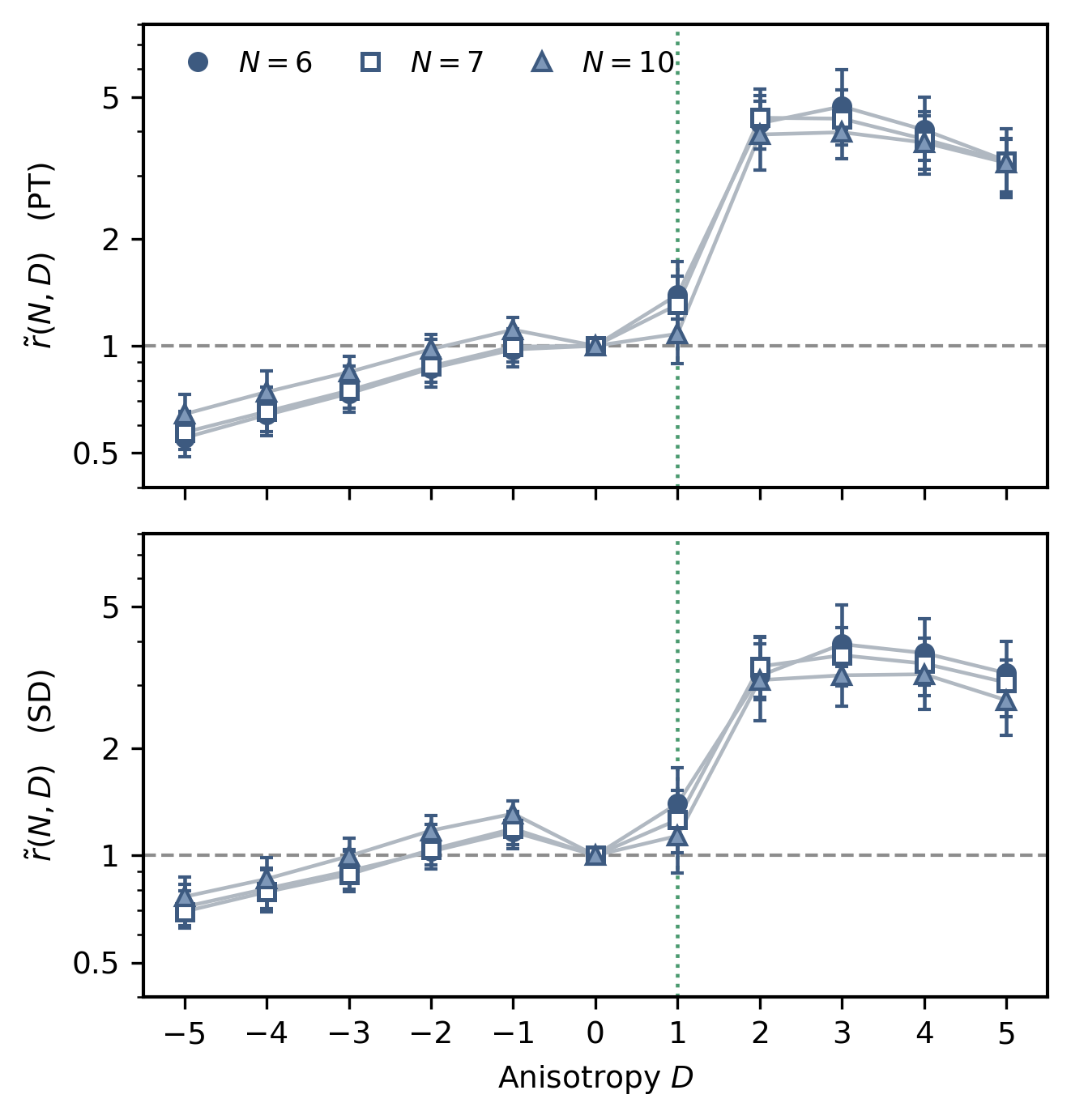}
\caption{Convention-free ratio $\tilde r(N,D)$ of Eq.~\eqref{eq:rtilde}, the anisotropy response of a classical solver normalized by that of the annealer on the identical instances. Top: parallel tempering. Bottom: random-restart steepest descent. Values above unity indicate that the classical solver degrades relative to the annealer as $D$ moves away from zero. Error bars are $95\%$ percentile bootstrap intervals with numerator and denominator resampled jointly. Dashed line: $\tilde r=1$; dotted line: $D=1$. The cost-conversion constant $\kappa$ cancels identically in $\tilde r$.}
\label{fig:rtilde}
\end{figure}

Figure~\ref{fig:rtilde} makes that separation quantitative in the only cross-protocol form free of a cost-conversion convention. The ratio $\tilde r$ rises monotonically along the axial sector, from about $0.6$ at $D=-5$ through unity at $D=0$ by construction, to $1.39$ ($95\%$ CI $[1.19,1.72]$), $1.31$ ($[1.05,1.57]$) and $1.08$ ($[0.89,1.37]$) at $D=1$ for $N=6,7,10$: the annealer's anisotropy response is more favourable than the classical one throughout the sweep toward $D=1$, significantly so at the two smaller sizes, with the trend itself common to all three. Between $D=1$ and $D=2$, exactly across the region where the crossing structure of Fig.~\ref{fig:gapmin} reorganizes, $\tilde r$ jumps by a further factor of three to four and then saturates. The lower panel repeats the construction with steepest descent in place of parallel tempering and reproduces the same curve within the confidence intervals, so the effect is not a property of thermal sampling in particular but of local classical search on this landscape.

Taken together, the spectral, dynamical, and landscape diagnostics show that single-ion anisotropy reshapes the annealing dynamics through the intermediate-state structure it creates.

\section{Conclusion}\label{Con}

We have investigated how single-ion anisotropy modifies finite-time quantum annealing in spin-$1$ Ising chains. The central question was whether the intermediate $m_z=0$ level can serve not only as part of a ternary encoding, but also as an active dynamical resource during the anneal. Our results show that its energetic accessibility, controlled by $D$, qualitatively reshapes the instantaneous spectrum and changes the pathways followed by the evolving state.

Three complementary observations establish this. First, the avoided-crossing structure is reorganized along the anisotropy axis: the single mid-schedule crossing of the easy-axis regime collapses to a much smaller gap late in the schedule near $D\approx1$, resolving into two distinct crossings at $N=7$ and $N=10$, while interior crossings disappear altogether in the deep easy-plane regime. Second, the mean occupation of the $m_z=0$ level witnesses transport through intermediate configurations, consistent with reconfiguration proceeding through a sequence of smaller transitions rather than a single collective rearrangement. Third, the finite-time loss of ground-state population is concentrated near the anisotropy-dependent crossings, linking the spectral restructuring directly to the dynamics. Since the anisotropy term acts as a constant on the $m_z=\pm1$ sector, every $D$ dependence reported here is necessarily mediated by amplitude in the intermediate level.

Because $D$ also enters the classical cost function as an on-site localization term, the anisotropy sweep is a sweep over a family of ternary optimization problems rather than a retuning of a single fixed one. Our comparison is therefore difficulty-controlled by construction: at every anisotropy the quantum protocol and the classical solvers are evaluated on the same instances, with the same exactly known ground states, and the anisotropy dependence of each is quantified relative to that solver's own performance at $D=0$.

The discriminating result is the behavior near $D\approx1$, where the quantum and classical responses separate. Relative to $D=0$ the annealer's optimized time-to-solution decreases significantly at $N=6$ and is statistically unchanged at $N=7$ and $N=10$, while both classical solvers slow down, parallel tempering significantly at the two larger sizes and steepest descent at all three, so the convention-free ratio $\tilde r$ exceeds unity at every size and significantly at $N=6$ and $N=7$. The exact landscape analysis of Appendix~\ref{App:Ruggedness} shows that the median escape barrier falls by roughly a factor of three between $D=0$ and $D=1$: energetic trapping weakens exactly where the classical dynamics slows. What changes instead is the multiplicity of near-degenerate configurations with mixed $m_z=0$ occupation, the same manifold presenting a degenerate region to a local classical search and furnishing the intermediate states through which the coherent dynamics proceeds. That two structurally unrelated classical dynamics, one thermal and one greedy, move in the same direction while the annealer does not cannot be attributed to a change in problem difficulty alone, and is independent of any convention relating annealing time to Monte Carlo work. Several limitations bound this conclusion. We claim no quantum speedup and no asymptotic advantage: the instances are one-dimensional chains with finite-range interactions, exactly solvable by dynamic programming and chosen deliberately as controlled testbeds. The accessible sizes are small and not contiguous, so we report size dependence without extrapolation, noting that the collapsed transition resolves into two distinct crossings only at $N=7$ and $N=10$, and that the ensemble separation at $D\approx1$ is not significant at $N=10$. The cross-protocol comparison is operational, its native cost units being incommensurable, so only its $D$ dependence is convention-free. Finally, the simulations are closed-system: decoherence and control noise are not modeled, the small gaps near $D\approx1$ may render that regime particularly sensitive to them, and the degeneracy-based reading of the classical slowdown rests on a structural correlate of the landscape rather than a first-principles bound on the search dynamics.

The present results motivate several natural extensions. A first direction is to test whether the intermediate-state mechanism persists in frustrated, higher-dimensional, or long-range ternary problems, where the underlying optimization landscape is no longer classically simple. Larger and more contiguous system sizes could be explored using tensor-network or Krylov-based time evolution to track how the anisotropy-induced crossing structure evolves with size. The same mechanism may also extend to higher-spin and qudit architectures, where multiple intermediate levels could support richer stepwise pathways through configuration space. Finally, platforms with tunable quadratic terms, including ultracold atoms, trapped-ion qudits, and engineered multilevel superconducting circuits, provide natural settings in which the robustness of anisotropy-controlled intermediate-state transport could be investigated experimentally.

\section*{Acknowledgments}
We thank \c{S}eyda Leyla Bozan for suggesting useful and relevant references.

\section*{Data availability statement}
The data that support the findings of this study are openly available at \href{https://github.com/coffee34900/Spin-1-quantum-annealing-with-anisotropy-controlled-intermediate-state-pathways}{REPO LINK}. The repository contains the simulation code, the disorder realizations used for all ensembles, and the scripts that reproduce the figures.

\appendix

\section{Energy Landscape Structure}\label{App:Ruggedness}

This appendix characterizes the classical energy landscape of the trit cost function $H(s)$, Eq.~\eqref{eq:classical_cost}, as a function of $D$, in order to interpret the $D$ dependence of the classical baselines observed in Sec.~\ref{Results}. All quantities are computed by exact enumeration of $\mathcal{S}=\{-1,0,+1\}^{N}$ for the identical disorder ensembles; clean chains are excluded because their exact symmetries produce degenerate plateaus that render the definitions below ambiguous.

The structure is probed with respect to the elementary single-site move of Sec.~\ref{tf}. Two configurations are neighbors if they differ at exactly one site,
\begin{align}
\mathcal{N}(s) = \Bigl\{ s' \in \mathcal{S} :
\sum_{i=1}^{N} \bigl(1 - \delta_{s_i, s_i'}\bigr) = 1 \Bigr\},
\qquad \abs{\mathcal{N}(s)} = 2N ,
\label{eq:neighborhood}
\end{align}
and the ground space at anisotropy $D$ is
\begin{align}
\mathcal{G} = \bigl\{ s \in \mathcal{S} : H(s) \le E_0 + \varepsilon \bigr\},
\qquad E_0 = \min_{s \in \mathcal{S}} H(s),
\label{eq:groundspace}
\end{align}
with the same tolerance $\varepsilon = 10^{-9}$ used for the success criteria. Four diagnostics are
considered.

First, a configuration is a strict local minimum if no neighbor has strictly lower energy, and we report the ensemble mean of their number,
\begin{align}
\mathcal{M} &= \bigl\{ s \in \mathcal{S} : H(s') \ge H(s) \;\; \forall\, s' \in \mathcal{N}(s)
\bigr\},
&
n_{\min} &= \abs{\mathcal{M}} .
\label{eq:localmin}
\end{align}
With continuous random couplings and fields exact ties occur with probability zero, so strict and weak minima coincide almost surely.

Second, greedy descent is described by the map
\begin{align}
g(s) = \operatorname*{arg min}_{s' \in \mathcal{N}(s) \cup \{s\}} H(s'),
\label{eq:greedymap}
\end{align}
which is almost surely single valued and satisfies $g(s)=s$ exactly on $\mathcal{M}$. Iterating $g$ from any configuration reaches a fixed point after finitely many steps, defining the basin map $b(s) = \lim_{k\to\infty} g^{k}(s) \in \mathcal{M}$ and the ground-basin fraction
\begin{align}
f_{gs} = \frac{1}{3^{N}} \abs{\bigl\{ s \in \mathcal{S} : b(s) \in \mathcal{G} \bigr\}},
\label{eq:fgs}
\end{align}
which measures how ``funnel-like'' the landscape is under purely local greedy moves. The same computation returns the mean number $w_{\rm desc}$ of single-site energy evaluations per descent, setting the cost unit of Eq.~\eqref{eq:tts_greedy}.

Third, for a local minimum $m \in \mathcal{M} \setminus \mathcal{G}$ we define the exact escape barrier through the connectivity of energy sublevel sets. Let $\Gamma_{m}$ denote the paths $\omega = (s^{(0)}{=}m, s^{(1)}, \dots, s^{(L)})$ with consecutive configurations neighbors in the
sense of Eq.~\eqref{eq:neighborhood} and endpoint energy $H(s^{(L)}) < H(m)$. Then
\begin{align}
E_{saddle}(m) &= \min_{\omega \in \Gamma_{m}}\; \max_{0 \le k \le L} H\bigl(s^{(k)}\bigr),\\
\Delta E(m) &= E_{saddle}(m) - H(m),
\label{eq:barrier}
\end{align}
the minimax energy over all single-site-move paths from $m$ to any strictly lower configuration. $\Delta E(m)$ controls thermally activated escape of a single Metropolis chain, with characteristic escape time $\sim e^{\beta \Delta E}$, and is evaluated exactly by ascending-energy flooding of the configuration graph: all $3^{N}$ configurations are processed in order of increasing energy and merged into connected clusters as they appear, and $E_{saddle}(m)$ is the energy at which the cluster containing $m$ first merges with a cluster containing a configuration of energy below $H(m)$. For each instance we take the median of $\Delta E$ over its non-ground minima and report the median of that quantity over the ensemble; instances with no non-ground minimum are excluded, and the number of contributing instances is stated wherever it is not close to $M$.

Fourth, as a measure of low-energy multiplicity complementary to the barrier diagnostics, we count the near-ground states within a window $\delta$ above the ground energy,
\begin{align}
n_{low} = \abs{\bigl\{ s \in \mathcal{S} : H(s) \le E_0 + \delta \bigr\}},
\qquad \delta = 1,
\label{eq:nlow}
\end{align}
in units of the couplings; the picture below is unchanged for $\delta = 0.5$.

\begin{figure}[tb]
\includegraphics[width=\columnwidth]{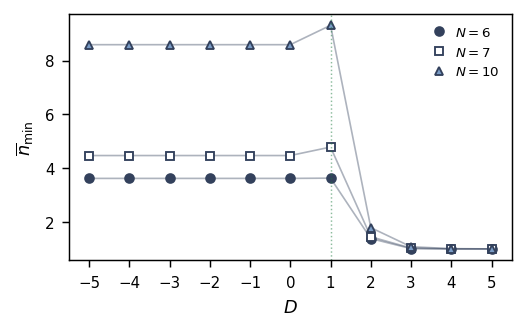}
\caption{Mean number of strict local minima $\overline{n}_{\min}$, Eq.~\eqref{eq:localmin}, versus anisotropy $D$, for the disorder ensembles of Sec.~\ref{Methods}. Dotted vertical line: $D=1$.}
\label{fig:lm}
\end{figure}

\begin{figure}[tb]
\includegraphics[width=\columnwidth]{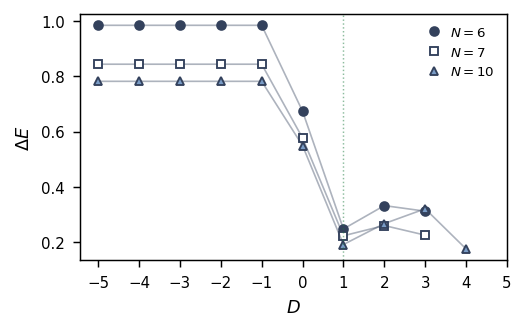}
\caption{Median escape barrier $\Delta E$, Eq.~\eqref{eq:barrier}, over the non-ground local minima of the ensemble, computed by exact ascending-energy flooding of the configuration graph. The barrier drops sharply through $D=1$ (dotted line), excluding an energetic origin for the classical slowdown observed there; points are absent where no non-ground minima occur. For $D\ge2$ the statistic is carried by the small and rapidly shrinking subset of instances that still possess a non-ground minimum (see text) and is not directly comparable with the values at $D\le1$.}
\label{fig:barrier}
\end{figure}

\begin{figure}[tb]
\includegraphics[width=\columnwidth]{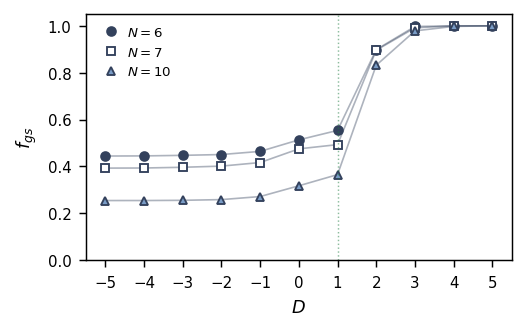}
\caption{Ground-basin fraction $f_{gs}$, Eq.~\eqref{eq:fgs}: the fraction of all $3^{N}$ configurations whose steepest-descent path terminates in the exact ground space. The fraction decreases with system size in the axial sector, and the landscape becomes nearly funnel-like for $D\ge 2$.}
\label{fig:fgs}
\end{figure}

\begin{figure}[tb]
\includegraphics[width=\columnwidth]{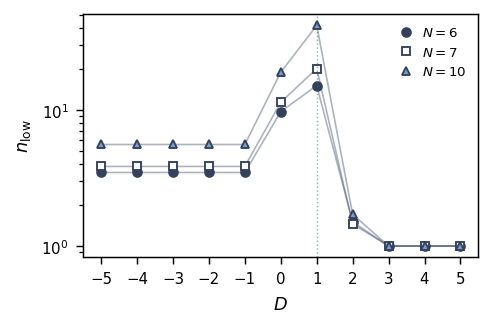}
\caption{Mean number of near-ground states $n_{low}$, Eq.~\eqref{eq:nlow}, within the window $E_0+1$ (logarithmic scale). The sharp peak at $D=1$ (dotted line), growing with system size, identifies the near-degeneracy of the low-energy spectrum discussed in the text.}
\label{fig:nlow}
\end{figure}

Figures~\ref{fig:lm}--\ref{fig:nlow} display the four diagnostics. For $D \le -1$ all structural measures are essentially independent of $D$: the low-energy configurations are composed almost entirely of $s_i = \pm 1$, on which the anisotropy term is the constant $DN$, so varying $D$ in this sector rigidly shifts the relevant part of the spectrum without reorganizing it. In this regime the mean number of local minima grows with system size from $3.5$ at $N=6$ through $4.3$ at $N=7$ to $8.4$ at $N=10$, the median escape barrier is of order the coupling scale ($0.70$--$1.11$), and only $26$--$46\%$ of configuration space drains into the ground basin under greedy descent, a fraction that decreases with system size. Already at $D=0$ the barrier begins to soften (median $0.50$--$0.76$), signalling the approach of the degeneracy region.

For $D \ge 3$ the landscape trivializes: the $m_z=0$ sector dominates, the mean number of local minima drops to essentially one, the ground state itself ($1.02$--$1.04$); non-ground minima survive in only $4$, $5$ and $6$ of $200$ instances at $N=6,7,10$, and vanish entirely at the largest anisotropies, where the barrier median is undefined; and the ground-basin fraction is indistinguishable from unity ($0.99$). The crossover regime $D=2$ is already strongly funnel-like ($f_{gs}=0.85$--$0.90$). This simplification must be kept in mind when interpreting the easy-plane sector in Sec.~\ref{Results}: part of the improved finite-time behavior of all three solvers there reflects properties of the problem itself, which is precisely why the main text characterizes the quantum mechanism through its full $D$ dependence and through the dynamical diagnostics, rather than through performance at any single anisotropy value.

The intermediate region $D \approx 1$ is where the landscape picture becomes discriminating. A natural first hypothesis for the classical slowdown observed there  (median time-to-solution larger than at $D=0$ by factors $1.02$, $1.24$, $1.12$ for parallel tempering and $1.02$, $1.20$, $1.18$ for steepest descent at $N=6,7,10$) would be enhanced energetic trapping: more local minima or higher barriers. The landscape diagnostics do not support this interpretation. The number of local minima changes only marginally at $D=1$ ($3.5\to3.6$ at $N=6$, $4.3\to4.7$ at $N=7$, $8.4\to9.3$ at $N=10$), while the median escape barrier drops by roughly a factor of three through this point, from $0.50$--$0.76$ at $D=0$ to $0.18$--$0.25$ at $D=1$. Energetic trapping becomes weaker, not stronger, exactly where the classical dynamics slows down. Beyond $D=2$ the barrier statistic is carried by a handful of instances and we read no trend into it.

What changes dramatically at $D \approx 1$ is instead the low-energy degeneracy (Fig.~\ref{fig:nlow}). The mean $n_{low}$ is sharply peaked there: at $N=10$ it reaches $46$, against $20$ at $D=0$, $5.5$ in the axial sector, and $1$ for $D \ge 3$; the enhancement relative to $D=0$ grows with system size ($1.4$, $2.1$, $2.3$ for $N=6,7,10$). The origin is transparent: near $D \approx 1$ the single-ion term makes local $m_z=0$ occupation energetically competitive with $m_z=\pm1$ without letting either sector win, so the spectrum just above the ground state is flooded with near-degenerate configurations differing by individual $\pm1 \leftrightarrow 0$ substitutions. The resulting landscape has low barriers but a nearly flat, highly degenerate floor, over which a local classical search must wander before committing to the ground state. The slowdown at $D \approx 1$ is thus consistent with an energetic mechanism driven by low-energy multiplicity rather than an energetic mechanism, and the fact that a greedy dynamics with no temperature at all slows in the same way and by a similar factor supports that reading. These results associate the classical slowdown with enhanced low-energy multiplicity rather than with increased escape barriers.

\section{Convention-dependent cost ratio}\label{App:Ratio}

The direct per-instance cost ratio is,
\begin{align}
r(N,D) = \kappa\,
\frac{TTS^{opt}_{PT}(N,D)}{TTS^{opt}_{AQA}(N,D)},
\label{eq:tts_ratio}
\end{align}
where $\kappa$ is a declared cost-conversion constant relating one unit of annealing time to one elementary single-site update. On a logarithmic scale, changing $\kappa$ shifts $r(N,D)$ uniformly without altering its shape, so its absolute scale depends on the declared value and only the location of the break-even point $r=1$ moves with it, whereas its dependence on $D$ is captured convention-independently by $\tilde r$ of Eq.~\eqref{eq:rtilde} in Sec.~\ref{Results}.

\begin{figure}[!htb]
\includegraphics[width=\columnwidth]{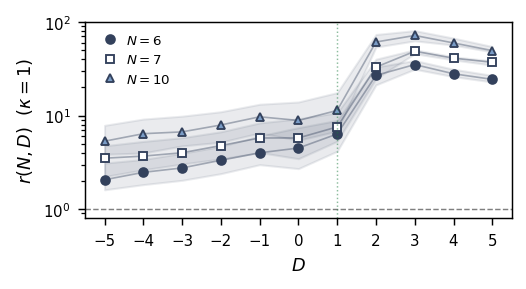}
\caption{Median per-instance cost ratio $r(N,D)$ of Eq.~\eqref{eq:tts_ratio} over the disordered ensembles, for the declared conversion $\kappa=1$, i.e.\ one unit of annealing time counted as one elementary single-site update. Shaded bands are $95\%$ percentile bootstrap intervals on the median. Dashed line: $r=1$; dotted line: $D=1$. The absolute scale is a convention; only the shape in $D$ is meaningful, and that shape is reported convention-free in Fig.~\ref{fig:rtilde}.}
\label{fig:ratio}
\end{figure}

Figure~\ref{fig:ratio} shows the median per-instance $r$ for the illustrative choice $\kappa=1$. With the parallel-tempering baseline tuned over both the sweep budget and the replica count, $r$ ranges from $1.1$ at $N=6$, $D=-5$, close to break-even under this convention, to $16$ at $N=10$, $D=3$. The curves shift upward with $N$ at fixed $\kappa$; we note this trend without extrapolating beyond the studied sizes, and caution that shifting $\kappa$ by one order of magnitude would move the entire family across the break-even line.
\bibliography{QA}

\end{document}